\documentclass[12pt]{article}

\usepackage{scicite}

\usepackage{times}

\usepackage[dvips]{graphicx}%

\newenvironment{sciabstract}{%
\begin{quote} \bf}
{\end{quote}}

\newcounter{lastnote}
\newenvironment{scilastnote}{%
\setcounter{lastnote}{\value{enumiv}}%
\addtocounter{lastnote}{+1}%
\begin{list}%
{\arabic{lastnote}.}
{\setlength{\leftmargin}{.22in}}
{\setlength{\labelsep}{.5em}}}
{\end{list}}

\title{Solid or Liquid ?  - Kinetically induced phase transition of a confined liquid}

\author
{Shivprasad Patil,$^{1}$ George Matei,$^{1}$ Ahmet Oral,$^{2}$ Peter M.\ Hoffmann $^{1\ast}$\\
\\
\normalsize{$^{1}$Department of Physics and Astronomy, Wayne State University,}\\
\normalsize{666 W Hancock, Detroit, MI 48201, USA}\\
\normalsize{$^{2}$Department of Physics, Bilkent University, 06800 Ankara, Turkey}\\
\\
\normalsize{$^\ast$To whom correspondence should be addressed;
E-mail:  hoffmann@wayne.edu.} }

\date{}

\begin{document}

\baselineskip24pt

\maketitle

\begin{sciabstract}
There has been long-standing debate about the physical state and
possible phase transformations of confined liquids. In this report
we show that a model confined liquid can behave both as a
Newtonian liquid with very little change in its dynamics or as a
pseudo-solid depending solely on the {\it rate} of approach of the
confining surfaces. Thus, the confined liquid does {\it not}
exhibit any confinement induced solidification in thermodynamic
equilibrium. Instead, solidification is induced kinetically, when
the two confining surfaces are approached with a minimum critical
rate. This critical rate is surprisingly slow, of the order of 6
\AA /s, explaining the frequent observation of confinement
induced solidification.
\end{sciabstract}

The structure and dynamics of confined liquids is of great
importance in interfacial phenomena from cell membranes to
nanotribology \cite{bhushan95}. In nanoscale confined liquids,
continuum theories break down and geometrically induced molecular
layering is observed \cite{israel92,yu99,donnelly02}. New tools to
study confined liquids include Surface Force Apparatus
(SFA)\cite{israel88,granick91}, Atomic Force Microscopy
(AFM)\cite{oshea94} and spectroscopic
techniques\cite{heuberger01}, such as Fluorescence Correlation
Spectroscopy (FCS)\cite{ashis02}. Various experiments have
yielded mutually exclusive findings on the {\it dynamics} of these
confined systems. In OMCTS (octamethylcyclotetrasiloxane), a
non-polar, roughly spherical, 'model' liquid, different research
groups have reported behavior ranging from crystallization, glass
formation, to no transition at all \cite{kumacheva,demirel}. The
same is true for water, the primary biological solvent
\cite{xhu01,raviv01,jeffery04}. Here, we report on recent
measurements of confined OMCTS using AFM. By systematically
altering measurement conditions, we found that dynamical
properties change profoundly as a result of a small change in the
compression rate. Thus the observation of confinement induced
solidification may be due to a kinetically induced transition
from liquid to solid.

\vspace{12 pt}

Measurements were performed with a home-built AFM that
incorporates a fiber interferometer to measure changes in the
cantilever's amplitude and phase \cite{patil05}. We vibrated the
AFM cantilever far below the resonance frequency and monitored the
amplitude and phase using a lock-in amplifier as the sample was
approached toward the tip. By using small cantilever amplitudes,
smaller than the size of a single molecule, we linearized the
measurement allowing us to directly relate the measured
cantilever phase and amplitude to the stiffness and damping
coefficient of the confined liquid \cite{jeffery04}. The sample
consisted of OMCTS, sandwiched between a flat silicon dioxide
surface and the silicon AFM tip. The OMCTS was purified by
passing it through molecular sieves and filtering it through a 20
nm filter just prior to each measurement. The silicon oxide
surfaces were prepared by oxidation in a heated Piranha solution
(1:3 H$_2$O$_2$, H$_2$SO$_4$) and drying in an oven at 120
$^\circ$C overnight. Experiments were performed at room
temperature (25 $^\circ$C).

To model the dynamic behavior of the liquid we found it
convenient to use the simplest viscoelastic model for a liquid,
the Maxwell model \cite{find}. It consists of a linear spring and
linear viscous element in series and exhibits time-dependent
stress dissipation under application of an external strain. The
characteristic relaxation time is given by $ t_R = k/(\gamma
\omega^2)$, where $k$ is the measured junction stiffness,
$\gamma$ is the measured damping coefficient and $\omega$ is the
oscillation (angular) frequency of the cantilever. In a liquid,
stresses will dissipate quickly and $t_R$ is expected to be low.
In ideal (elastic) solids, stresses can be sustained
indefinitely, so $t_R$ is expected to be large if the system
behaves more solid-like.

\vspace{12 pt}

Fig.\ 1a shows the stiffness and normalized damping coefficient
measured at an approach rate of 3 \AA /s. The damping coefficient
is normalized to the measured bulk value far from the surface.
This value varied between different levers, but was generally in
the range of $10^{-5}$ Ns/m. It can be seen that the stiffness
and damping coefficient are 'in-phase', i.\ e.\ maxima of the
stiffness are aligned with maxima in the damping coefficient.
Since a higher stiffness implies an increased density of the
confined liquid, the confined liquid acts like a Newtonian fluid,
where the viscosity is expected to increase with density. Fig.\
1b shows the normalized mechanical relaxation time of the liquid,
$t_R$, calculated from the data in Fig.\ 1a. Again the relaxation
time is normalized to the measured bulk value, which typically was
of order $2 \times 10^{-4}$ seconds. The mechanical relaxation
time does not show any systematic changes associated with either
confinement or layering. Thus, the dynamics of the system seems
to be unaffected by confinement and the liquid remains
liquid-like even at small separations. The only change that is
induced is a density oscillation as a function of separation
associated with the geometrically induced layering of the liquid
molecules.

\begin{figure}[h]
\begin{center}
$\begin{array}{c@{\hspace{10mm}}c} \multicolumn{1}{l}{\mbox{\bf
(a)}} &
     \multicolumn{1}{l}{\mbox{\bf (b)}} \\ [-0.53cm]
\includegraphics[width=65mm, clip,
keepaspectratio]{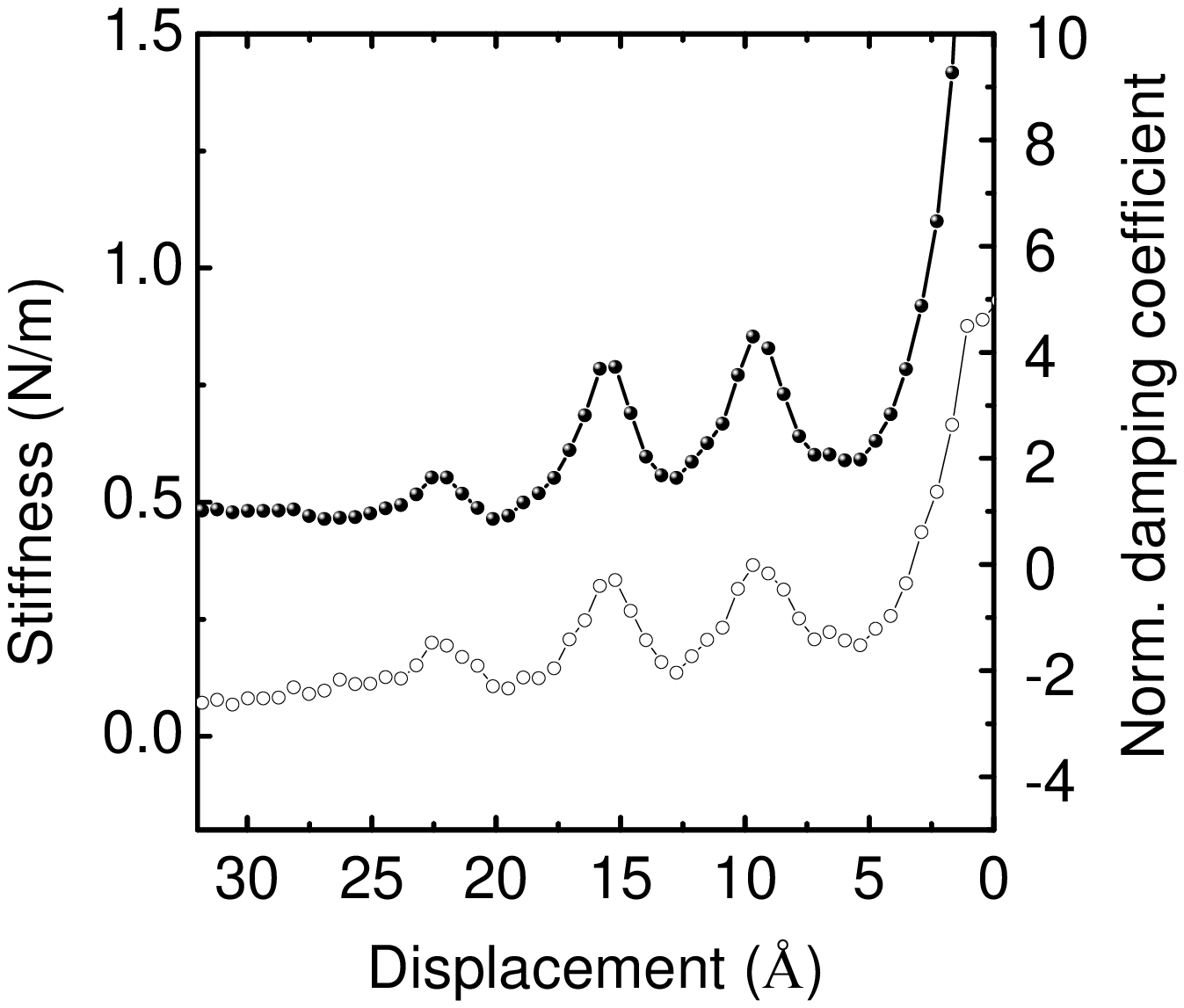} &
\includegraphics[width=65mm, clip,
keepaspectratio]{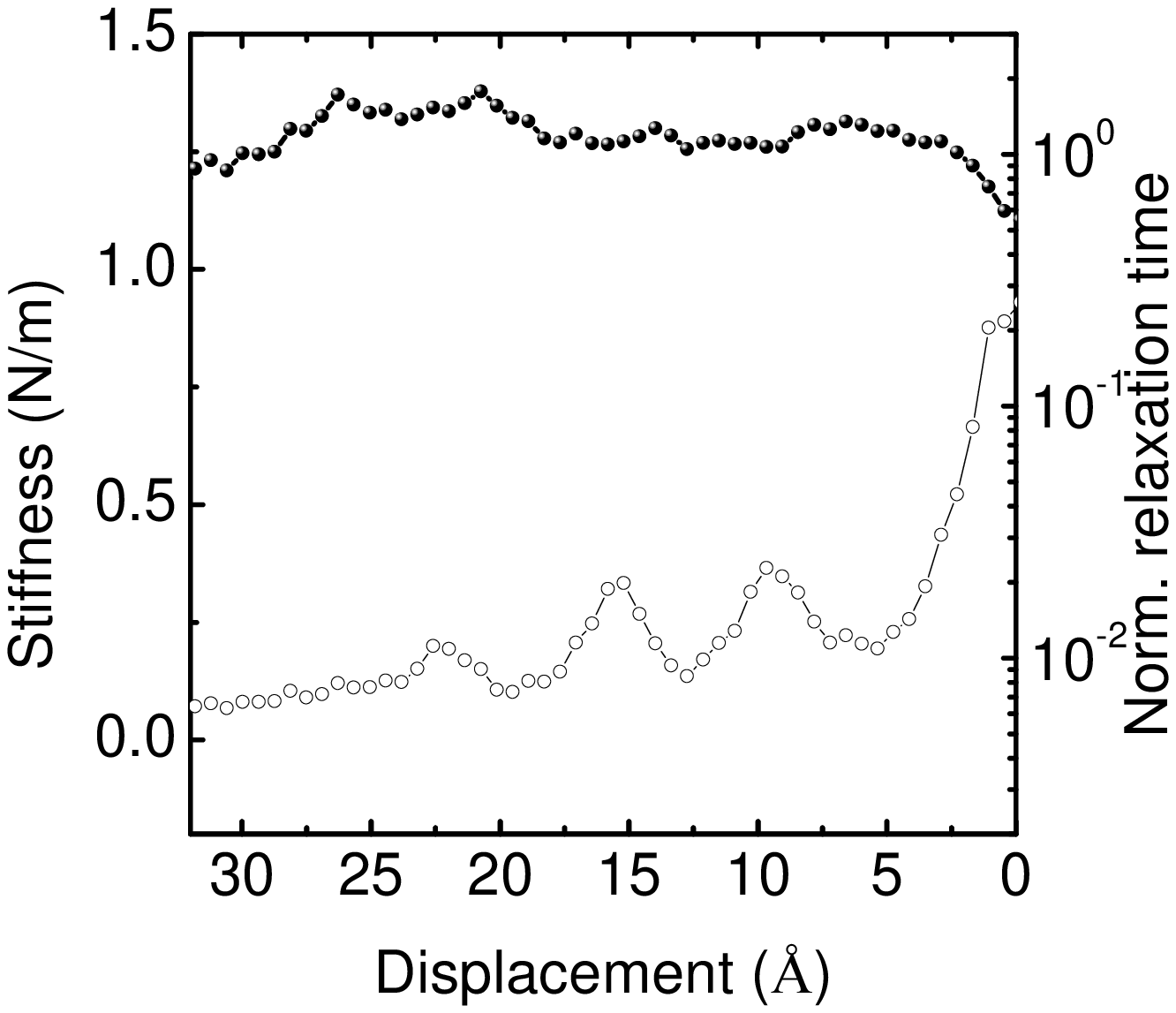}
\end{array}$
\end{center}
\caption{(a) Junction stiffness (open circles) and normalized
damping coefficient (line graph) versus displacement for OMCTS
confined between the AFM cantilever tip and a silicon oxide
surface. Cantilever frequency and free amplitude were 460 Hz and
3.5 \AA, respectively.  The sample was approached towards the tip
at a 'slow' rate of 3 \AA/s. Clear stiffness oscillations can be
seen with an average separation of about 9 \AA, consistent with
the diameter of OMCTS molecules. The stiffness and the damping
coefficient are in-phase in this case. (b) Junction stiffness
(open circles) and Maxwell mechanical relaxation time (filled
circles) versus displacement. Note that the relaxation time does
not systematically change with displacement.}
\end{figure}

By contrast, Fig.\ 2a shows stiffness and damping at an approach
rate of 12 \AA /s. We can see a dramatic change in the relative
magnitude of stiffness and damping coefficient. They are now
'out-of-phase', and the liquid shows {\it reduced} damping in the
'ordered', high stiffness state and liquid-like damping in the
disordered, low stiffness state. This suggests that in the ordered
state the liquid now behaves more like an elastic solid. Fig.\ 2b
shows the corresponding relaxation time which now shows clear
oscillations, with prominent maxima associated with the 'solid',
ordered state of the liquid.

\begin{figure}[h]
\begin{center}
$\begin{array}{c@{\hspace{10mm}}c} \multicolumn{1}{l}{\mbox{\bf
(a)}} &
     \multicolumn{1}{l}{\mbox{\bf (b)}} \\ [-0.53cm]
\includegraphics[width=65mm, clip,
keepaspectratio]{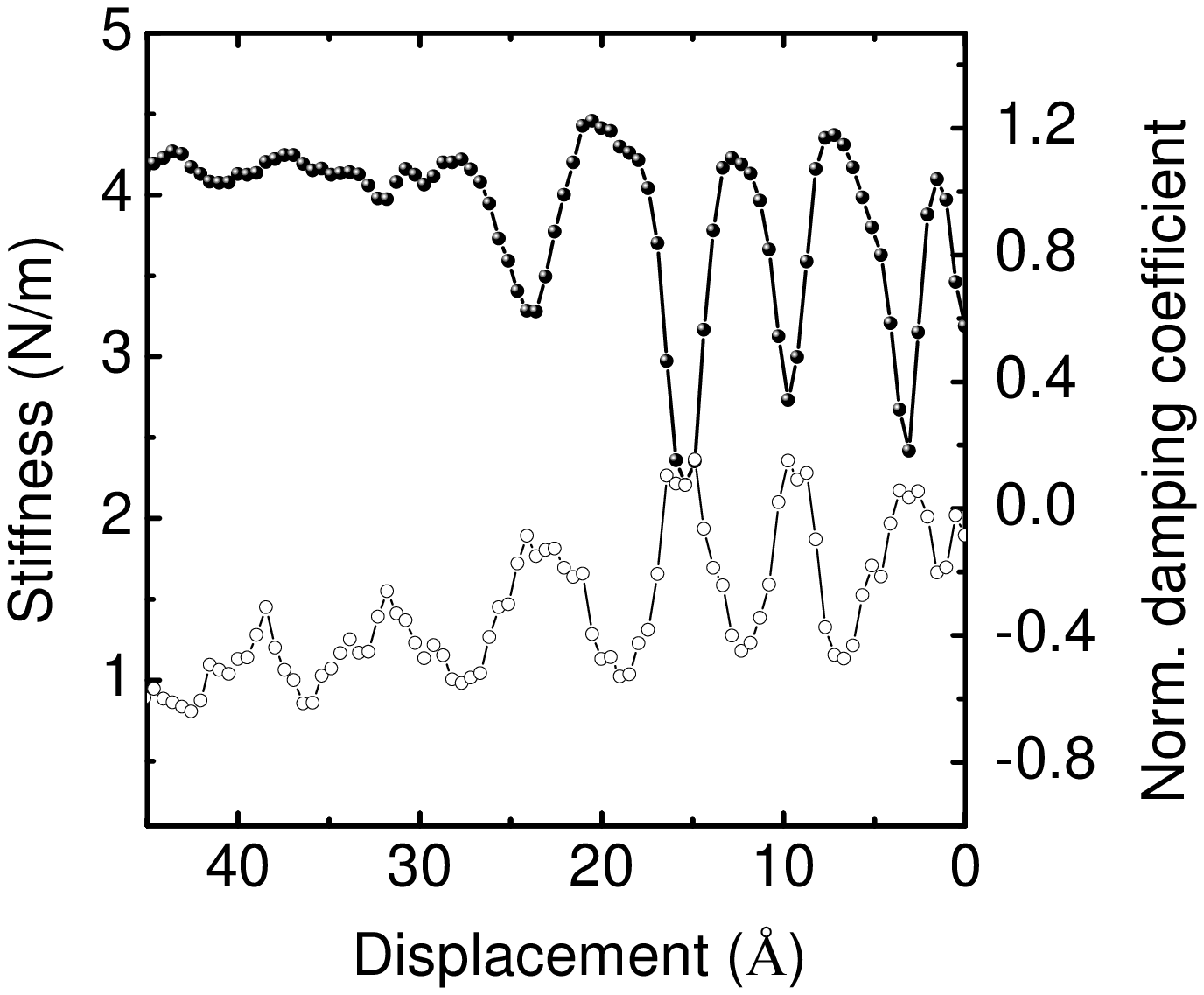} &
\includegraphics[width=65mm, clip,
keepaspectratio]{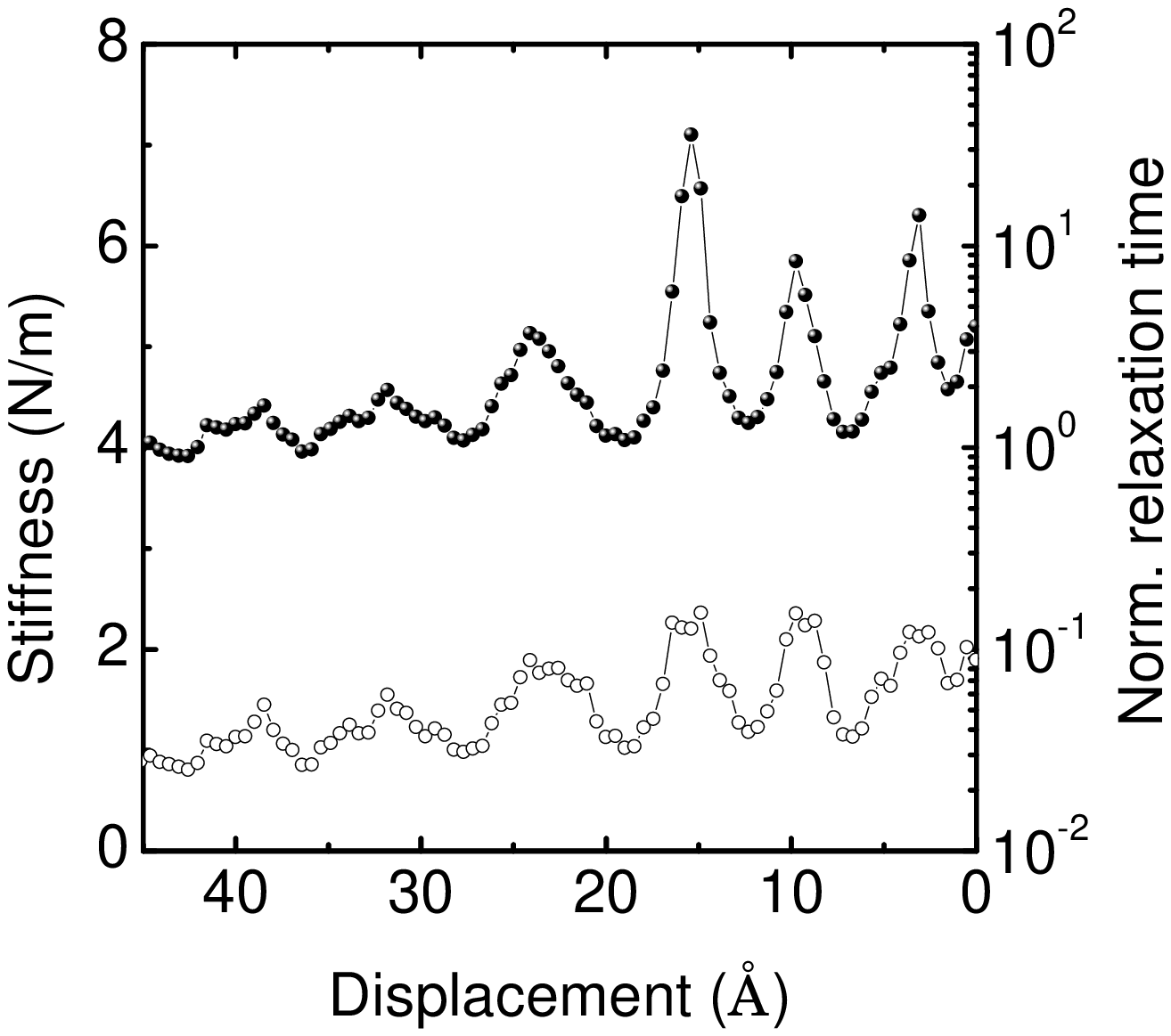}
\end{array}$
\end{center}
\caption{(a) Junction stiffness (open circles) and normalized
damping coefficient (line graph) versus displacement for OMCTS
confined between the AFM cantilever tip and a silicon oxide
surface. Cantilever frequency and free amplitude were 400 Hz and
2.4 \AA, respectively. In this case, the sample was approached
towards the tip at a 'fast' rate of 12 \AA/s. Clear stiffness
oscillations can be seen with an average separation of about 9
\AA, consistent with the diameter of OMCTS molecules. The
stiffness and the damping coefficient are now out-of-phase with
respect to each other. (b) Junction stiffness (open circles) and
Maxwell mechanical relaxation time (filled circles) versus
displacement. Note that the relaxation time shows strong peaks
associated with the high stiffness regions of the sample.}
\end{figure}

Fig.\ 3 summarizes our measurements. We plotted our observations
in a matrix of two crucial parameters: The oscillation frequency
of the cantilever and the approach rate. In our experiments, we
also explored different cantilever amplitudes in a restricted
range from 1.5 \AA to 7 \AA, but except for an attenuation of the
peak heights in the stiffness at larger amplitudes, we did not
find any systematic dependence in this range of small amplitudes.
As can be seen in Fig.\ 3, the approach rate, rather than the
oscillation frequency of the lever, is the crucial parameter that
determines the dynamical behavior of the liquid. This may seem
surprising since the maximum speed of the cantilever during each
oscillation cycle is of the order of 1000 \AA /s, i.\ e.\ much
larger than the approach speed. However, the lever is oscillated
at small amplitudes compressing the liquid film only slightly
without squeeze-out of a complete molecular layer. Thus, the
oscillation of the lever probes the mechanical and dynamical
properties of the confined liquid film, and the slow approach of
the tip actually forces molecular layers to be pushed out of the
tip-sample gap.  We can thus postulate that the observed kinetic
phase transformation may be due to the fact that at a rate of 6
\AA/sec and above, the molecular layers are 'jammed' and are not
able to react to the narrowing of the gap fast enough.

\begin{figure}\centerline{\includegraphics[width=86mm, clip,
keepaspectratio]{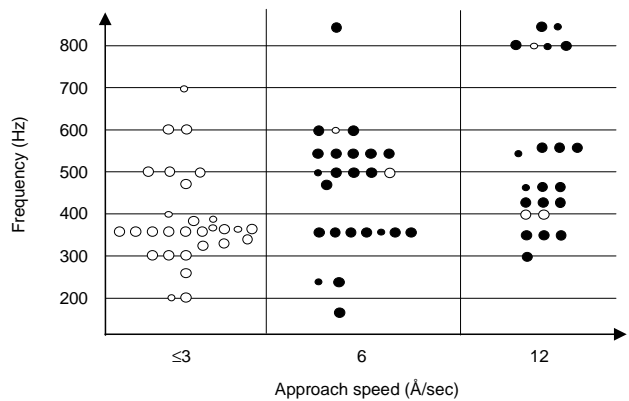}} \caption{Summary of all measurements that
showed stiffness oscillations plotted in a matrix of approach
speed and cantilever oscillation frequency. Due to local
roughness of the sample, not all measurements showed stiffness
oscillations. The typical 'success' rate in our measurements was
of the order of 20 \%. Open circles denote cases in which
stiffness and damping were in-phase and the mechanical relaxation
time was essentially constant with displacement - as in Figure 1
(liquid-like behavior). Filled circles denote measurements where
stiffness and damping were out-of-phase and the mechanical
relaxation time showed distinct peaks associated with the high
stiffness phase of the confined fluid - as in Figure 2 (solid-like
behavior). It can be seen that liquid- or solid-like behavior
depends merely on the approach speed and not on the oscillation
frequency of the cantilever. Size of circles denotes confidence in
data - small circles denote cases where stiffness and damping
peaks were aligned either in-phase or out-of-phase, but data was
noisy or alignment of peaks changed as surface was approached (in
these cases the alignment closer to the surface was used). In all
other cases (large circles) the relative alignment of stiffness
and damping peaks was clearly either in or out-of phase over the
full range of observed peaks. }\label{fig1}
\end{figure}

It is useful to compare our results with recent experiments by Xhu
and Granick \cite{xhu03} who report the onset of strong friction
in confined OMCTS layers only if the layers were squeezed at a
rate exceeding 5 \AA/s. If the surfaces were approached much
slower, friction was immeasurably small. Thus, the confined liquid
became 'solidified' at approach rates of $\geq$ 5 \AA/s, almost
identical to our observations. If we define a critical time scale
by dividing the thickness of one molecular layer by the critical
approach rate, we find a critical time of the order of 9 \AA /(6
\AA /s) = 1.5 s, which is 12-14 orders of magnitude longer than
typical molecular relaxation times. The identical time-scale
observed in the SFA experiments by Xhu and Granick \cite{xhu03}
and our present observations - despite the vastly different
lateral dimensions of the confined region ($\approx 10 \mu$m
versus $\approx 10$ nm)- suggest that this behavior is
independent of the lateral size of the confined region. Rather it
must be intrinsic to characteristic properties of the confined
liquid and the thickness of the film (number of layers).

There are few models describing the squeeze-out dynamics of
confined liquids. Persson et al.\ \cite{persson94} proposed a
nucleation model. However, we found that the critical parameters
obtained from this theory (in particular the pressure) do not
match well with our experimental observations. Moreover, these
parameters do not change much if the approach rate is varied in
the range of 3 - 12 \AA /s because of the logarithmic dependence
of the critical values on the nucleation rate. Thus this approach
cannot explain the dramatic shift in behavior from 3 to 6 \AA /s
approach rate. How about the time to expel the layer after a
'hole' has nucleated ? In a recent measurement by Becker et al.\
\cite{becker03}, an OMCTS layer of 25 $\mu$m radius was expelled
in about 2 seconds. According to Persson et al.\
\cite{persson94}, the squeeze-out time is proportional to the
area of film that needs to be expelled. Thus, with a 100 nm
radius tip, a layer should be expelled in about 10 - 100 $\mu$s.
This is much too fast to be observed in our measurements.

The mechanical behavior of this simple system changes profoundly
from liquid to solid depending on an experimentally imposed
(macroscopically long) time scale. The system exhibits a sharp
kinetically induced transition in response to a rather small
change in this time scale. This change (a factor of two) is very
small indeed, if we consider thermodynamic arguments where rates
are typically exponentially dependent on activation energies (like
the nucleation model discussed above). This suggests that under
confinement, i.e. effectively in two dimensions, it is difficult
for the molecules to move out of the way of the approaching
surfaces except by a slow {\it cooperative} process, involving a
{\it characteristic number of molecules} (which may depend on the
specific interactions between the molecules). If the molecules
are not given enough time, they become 'jammed', and the system
is forced into a non-equilibrium 'solid' state, exhibiting high
friction (Xhu and Granick) and an elastic response to normal
pressure (present work).

How can we estimate the number of molecules that have to act
cooperatively in order to arrive at a characteristic time of a few
seconds for one complete squeeze-out? A typical molecular
relaxation time (i.e. the average time a molecules moves freely
before colliding with another molecule) is of the order of
$\tau_0=10^{-14}$ s. Now, in order to move out of a layer beneath
the tip, the molecules have to move away from a central point,
opening up a hole in the layer. Treating the problem as a
2-dimensional problem, we can see that the probability that a
molecule will move away from a central point, rather than towards
it, is approximately 1/2 (since 1/2 of all possible angles of
motion will point away from a central point). Thus the probability
that $N$ molecules will all move away from a central point at the
same time is of order $p_N=(1/2)^N$. Then the mean time for the
event of $N$ molecules moving away from a central point, opening
up a hole in the layer, to happen randomly is of the order of
$\tau_0/p_N$. For this time to be 1.5 seconds, $N$ would have to
be about 47 molecules. Thus it does not take a very large number
of cooperatively moving molecules to arrive at macroscopic times.

These observations are reminiscent of the jamming transition in
suspensions of colloidal particles, granular materials or glassy
systems near the glass transition\cite{israeloff}. In granular
materials, for instance, fast compaction leads to an elastic
response, while slow compaction speeds allow for a more plastic
response  and 'greater internal rearrangement' \cite{sanchez03}.
Comparing the present observations to glassy systems, we find
some similarities as well (indeed, 'solidified' OMCTS has been
described as a glassy solid): The role of cooperativity and the
rate dependence (cooling rate in the case of glass transitions).
However, there are some important differences: The temperature in
our experiments was constant, and the kinetic transition was
induced by the rate of volume reduction and not by cooling. A
more striking difference is that OMCTS is a very simple system,
consisting of small globular molecules, unlike the typical
glass-forming systems, such as polymers or heterogeneous mixtures
of several constituents.

What are the practical implications of these observations? In
macroscopic systems, even in the absence of a lubricant, friction
is dominated by confined layers of contaminant hydrocarbon and
water layers \cite{he99}. In such systems, lateral motions are
typically not well controlled on a molecular scale and any shear
motion will be accompanied by a normal motion far exceeding 3
\AA/s. Thus, we could expect that macroscopic friction is
partially due to molecular jamming of lubricant molecules.
However, lubricants are often  a mixture of different molecules.
Thus layering may be greatly disrupted.

In microscopic situation, including Nano-electromechanical
systems (NEMS), our findings may provide insight for the
management of frictional dissipation. As long as approach rates
can be kept very low, lateral friction could be kept very low as
well. This should be possible in systems that are approaching
molecular dimensions. Furthermore, the fact that under faster
approach rates the system behaves elastically may lead to designs
that exploit the confined lubricant as a 'smart liquid' to control
approach rates in small devices.

Looking further afield, the important observation that even
simple systems can alter their properties profoundly as a
function of time scale, and that the critical time scale can be
macroscopically long could have important implications in many
areas of nanoscience and molecular biology.

\bibliographystyle{Science}

\begin{thebibliography}{35}
\bibitem{bhushan95} B. Bhushan, J. N. Israelachvili and U.
Landman, {\it Nature} {\bf 374}, 607 (1995).
\bibitem{israel92} J. N. Israelachvili, {\it Intermolecular and
Surface Forces} (Academic Press, San Diego, CA, 1992), Ch. 13.
\bibitem{yu99} C.-J. Yu, G. Richter, A. Datta, M. K. Durbin and P.
Dutta, {\it Phys. Rev. Lett.} {\bf 82}, 2326 (1999).
\bibitem{donnelly02} S. E. Donnelly, R. C. Birtcher, C. W. Allen,
I. Morrison, K. Furuya, M. Song, K. Mitsuishi and U. Dahmen, {\it
Science} {\bf 296}, 507 (2002).
\bibitem{israel88} J. N.
Israelachvili and P. M. McGuiggan, {\it Science} {\bf 241}, 795
(1988).
\bibitem{granick91} S. Granick, {\it Science} {\bf 253}, 1374 (1991).
\bibitem{oshea94} S. J. O'Shea, M. E. Welland and J. B. Pethica,
{\it Chem. Phys. Lett.} {\bf 223}, 336 (1994).
\bibitem{heuberger01} M. Heuberger, M. Z\"ach and N. D. Spencer,
{\it Science} {\bf 292}, 905 (2001).
\bibitem{ashis02} A. Mukhopadhyay, J. Zhao, S. C. Bae and S.
Granick, {\it Phys. Rev. Lett.} {\bf 89}, 136103 (2002).
\bibitem{kumacheva} E. Kumacheva and J. Klein, {\it Science}
{\bf 269}, 5225 (1995); J. Klein and E. Kumacheva, {\it J. Chem.
Phys.} {\bf 108}, 6996 (1998).
\bibitem{demirel} A. L. Demirel and S. Granick, {\it Phys. Rev.
Lett.} {\bf 77}, 2261 (1996); A. L. Demirel and S. Granick, {\it
J. Chem. Phys} {\bf 1115}, 1498 (2001).
\bibitem{raviv01} U. Raviv,
P. Laurat and J. Klein, {\it Nature} {\bf 413}, 51 (2001).
\bibitem{xhu01} Y. Xhu and S. Granick, {\it Phys. Rev. Lett.} {\bf
87}, 096104 (2001).
\bibitem{jeffery04} S. Jeffery, P. M.
Hoffmann, J. B. Pethica, Ch. Ramanujan, H. \"O. \"Ozer, and A.
Oral, {\it Phys. Rev. B} {\bf 70}, 054114 (2004).
\bibitem{patil05} S. Patil, G. Matei, H. Dong,
P. M. Hoffmann, M. Karak\"ose and A. Oral, {\it Rev. Sci. Instr.}
(2005), accepted for publication.
\bibitem{find} W. N. Findley, J. S. Lai and K. Onaran,
{\it Creep and Relaxation of Nonlinear Viscoelastic Materials}
(Dover, New York, 1976), Ch. 5.
\bibitem{xhu03} Y. Xhu and S. Granick, {\it Langmuir} {\bf 19},
8148 (2003).
\bibitem{persson94} B. N. J. Persson and E. Tosatti, {\it Phys.
Rev. B} {\bf 50}, 5590 (1994).
\bibitem{becker03} Th. Becker and F. Mugele, {\it Phys. Rev.
Lett.} {\bf 91}, 166104 (2003).
\bibitem{israeloff} E. V. Russell
and N. E. Israeloff, {\it Nature} {\bf 408}, 695 (2000).
\bibitem{sanchez03} F. X. Sanchez-Castillo and J. Anwar, {\it
Chem. Mater.} {\bf 15}, 3417 (2003).
\bibitem{he99} G. He, M. H. M\"user and M. O. Robbins, {\it
Science} {\bf 284}, 1650 (1999).

\end{thebibliography}

\begin{scilastnote}
\item P.\ M.\ Hoffmann would like to acknowledge generous support
by NSF (MRI grant DBI-0321011, ANESA grant INT-0217789, and CAREER
grant DMR-0238943). A. Oral and P. M. Hoffmann would also like to
acknowledge support for international collaborative research
through NSF and the Turkish Government (TUBITAK, Grant No.
TBAG-U/50(102T117)).
\end{scilastnote}

\clearpage

\section*{Supporting Online Material}

\textbf{Instrument:}In our experiments, we use the dynamic mode
of a home-built Atomic Force Microscope(AFM)\cite{patil05}. We
impose a small oscillation on the AFM cantilever using a
piezo-electric element and then monitor amplitude and phase angle
between the piezo motion and the cantilever end. The deflection
of the cantilever is measured using a highly sensitive fiber
interferometer with a measured base noise level of 600
fm/$\sqrt{\rm{Hz}}$. The end of the fiber is coated with a 30\%
reflective thin film.  This allows for multiple reflections
between the mirrored cantilever back and the fiber end, before
the light couples back into the fiber. To optimally align the
fiber with respect to the cantilever we built a five
degree-of-freedom inertial positioner with $\approx 50$ nm step
size on which the the fiber is mounted. Once we achieve the
desired sensitivity the fiber is locked in that position using a
feedback loop. This allows us to use small amplitudes of an
angstrom or less, linearizing the measurement. Jump-to-contact is
avoided by using sufficiently stiff levers. The cantilever is
oscillated far below the resonance to avoid resonance enhancement
of the amplitude and simplify data interpretation. The junction
stiffness and the damping coefficient is then directly related to
amplitude and phase in the following way.
\begin{equation}
k = k_L(\frac{A_0}{A} \cos\phi -1)
\end{equation}
and
\begin{equation}
\gamma = - \frac{k_L A_0}{A \omega} \sin\phi
\end{equation}
where $A_0$ is the free amplitude of the cantilever, $A$ is the
measured amplitude during the approach, $k_L$ is the stiffness of
the cantilever, $\phi$ is the phase angle between the cantilever
drive and the cantilever end, and $\omega$ is the angular
frequency of the cantilever oscillation.

\textbf{Sample Preparation:} In AFM and SFA studies of confined
liquids ordering is only observed if the confining surfaces are
atomically smooth.  In AFM, the tip needs to be sharp, such that
the measurement is dominated by forces acting on the very end of
the tip. If the tip is blunt, it is also typically rough on a
molecular scale and due to 'destructive interference' of layering
in different regions, no layering is observed in the
measurements. Any roughness in the sample also disrupts layering.
In our studies the OMCTS is confined between a silicon oxide
substrate and a silicon tip. For producing smooth silicon oxide,
silicon wafers are immersed in a strong oxidizer (hydrogen
peroxide and sulfuric acid in the proportion 1:3) and heated for
20 minutes at $100^\circ$C for 20 minutes. The process of
simultaneous etching and oxidation makes the surface free from
any contaminants. The wafers are then rinsed in DI water and kept
in the oven overnight at $120^\circ$C  to remove any remaining
water from the surface.

\textbf{Humidity control:} The liquid under study, OMCTS, is known
to be hygroscopic. Before use, OMCTS is stored overnight in a
sealed bottle containing molecular sieves which absorb the water.
The OMCTS is then passed through a syringe filter of .02 nm pore
size. If humidity is high, OMCTS gets contaminated by atmospheric
water during the course of the experiment. In some of our
measurements, where the sample was exposed to the air for long
times, we saw a clear attractive background in the force profiles.
To avoid this we place the AFM in a de-humidified chamber which
is isolated from the rest of the room. A de-humidifier is used to
maintain the humidity level below 30 \% throughout the
experiments.

\end{document}